\newcommand{\nd}{Ni(C$_5$H$_{14}$N$_2$)$_2$N$_3$(PF$_6$)}
\newcommand{\nenp}{Ni(C$_2$H$_8$N$_2$)$_2$NO$_2$(ClO$_4$)}

\documentstyle[prb,aps,twocolumn]{revtex}
\begin{document}
\input{psfig.sty}
\draft

\twocolumn[\hsize\textwidth\columnwidth\hsize\csname
@twocolumnfalse\endcsname

\title{Haldane-gap excitations in the low-$H_c$ 1-dimensional quantum antiferromagnet \nd\ (NDMAP).}

\author{A. Zheludev,$^\dag$ Y. Chen,$^\ddag$ C. L. Broholm,$^{\ddag,\sharp}$ Z. Honda$^{\ast}$ and K. Katsumata.$^{\ast}$}
\address{$\dag$ Physics Department, Brookhaven National Laboratory, Upton, NY
11973-5000, USA.\\}

\address{$\ddag$ Department of Physics and
Astronomy, Johns Hopkins University, Baltimore, MD 21218, USA.}

\address{$\sharp$ NIST Center for Neutron Research, National Institute of
Standards and Technology, Gaithersburg, MD 20899, USA.}

\address{$\ast$ RIKEN (The Institute of Physical and Chemical Research), Wako, Saitama 351-0198, Japan.}

\date{\today}
\maketitle
\begin{abstract}
Inelastic neutron scattering on deuterated single-crystal samples
is used to study Haldane-gap excitations in the new $S=1$
one-dimensional quantum antiferromagnet \nd\ (NDMAP), that was
recently recognized as an ideal model system for high-field
studies. The Haldane gap energies $\Delta_x=0.42$~meV,
$\Delta_y=0.52$~meV and $\Delta_z=1.86$~meV, for excitations
polarized along the $a$, $b$, and $c$ crystallographic axes,
respectively, are directly measured. The dispersion perpendicular
to the chain axis $c$ is studied, and extremely weak inter-chain
coupling constants $J_y=1.8\cdot 10^{-3}$~meV and $J_x=3.5\cdot
10^{-4}$~meV, along the $a$ and $b$ axes, respectively, are
determined. The results are discussed in the context of future
experiments in high magnetic fields.
\end{abstract}


]

\narrowtext

\section{Introduction}
The unique properties of the one-dimensional (1D) integer-spin
Heisenberg antiferromagnet (HAF) have captivated the minds of
condensed matter physicists for the last 20 years. In total
defiance of the quasi-classical picture of magnetism, the ground
state in this system is a spin-singlet- a ``quantum spin liquid''
with only short-range (exponentially decaying) spin correlations.
The excitation spectrum is also rather unique, and, even in the
absense of any magnetic anisotropy, features a so-called Haldane
energy gap.\cite{Haldane} The wealth of theoretical and
experimental results accumulated to date provide a fairly complete
understanding of the physics involved, and few mysteries remain,
as far as the behavior of the idealized model is concerned.
Particularly revealing was the neutron scattering work done on
real quasi-1D $S=1$ HAF compounds such as
CsNiCl$_3$,\cite{CsNiCl3} NENP (\nenp),\cite{NENP} and
Y$_2$BaNiO$_5$ (Ref.~\onlinecite{YBANO}).

One of the few remaining unresolved issues is the behavior of
Haldane-gap antiferromagnets in high magnetic fields. An external
magnetic field splits the triplet of Haldane
excitations,\cite{Regnault92,NENP,Golinelli93,Katsumata89,Affleck91,Sakai91}
driving one of the modes to zero energy at some critical field
$H_c$. At this soft-mode quantum phase transition, static
long-range antiferromagnetic correlations appear. In essence, the
external field suppresses zero-point quantum spin fluctuations
that are responsible for the destruction of long-range order in
the spin-liquid state, and restores long-range spin coherency.
Despite the success of high-field bulk studies of several
material, it has been highly frustrating, that for most existing
systems the actual values of the critical fields are unobtainable
in neutron scattering experiments. As a result, the properties of
the high-field phase are still largely unknown. It is even
unclear, whether or not the high-field phase is commensurate, and
virtually nothing is known about the excitation spectrum.

\begin{figure}
\parbox[b]{3.4in}{
\psfig{file=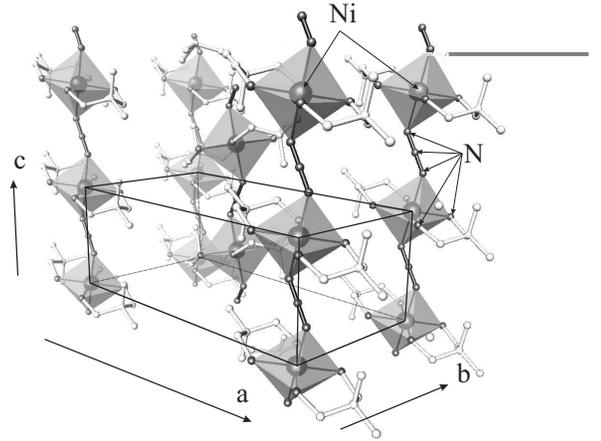,height=2.8in,angle=90} \caption{A schematic
view of the antiferromagnetic spin chains in the NDMAP crystal
structure. Only Ni, N and C atoms are shown.} \label{struct}
}\end{figure}

Traditionally, NENP was the workhorse of high-field studies. For
this compound large single crystal samples can be fabricated, and
the critical field $H_c=9$~T (when applied parallel to the
chain-axis) is, in principle, accessible in a neutron-scattering
experiment. In practice, it is however easier to perform
measurements with a magnetic field applied perpendicular to the
spin chains. The corresponding critical field for NENP is,
unfortunately, much larger: $H_c\approx 11-13$~T. In any case,
measurements deep within the high-field phase are not possible.
Moreover, due to certain structural features, the phase transition
in NENP is smeared out. The g-tensors of the $S=1$ Ni$^{2+}$ ions
are staggered in this compound and lead to an effective staggered
field when an external uniform field is
applied.\cite{Chiba91,Mitra94,Sakai94} As a result, a static
staggered magnetization appears at arbitrary weak applied fields,
and, strictly speaking, no additional spontaneous symmetry
breaking occurs at $H_c$.

The recent discovery of NDMAP (\nd),\cite{Monfort96} a relatively
easily crystallized  Haldane-gap compound with a critical field of
only around 4~T,\cite{Honda98} and no staggering of g-tensors
within spin chains, promises to make the high-field phase readily
accessible in neutron scattering studies. The crystal structure of
this material is similar to that of NENP, and is visualized in
Fig.~\ref{struct}. The AF spin chains run along the $c$ axis of
the tetragonal structure (space group $P_{nmn}$, $a=18.046$~\AA,
$b=8.705$~\AA, and $c=6.139$~\AA), and are composed of
octahedrally-coordinated $S=1$ Ni$^{2+}$ ions bridged by triplets
of nitrogen atoms. These long nitrogen bridges account for a
realtively small in-chain AF exchange constant $J=2.6$~meV, as
estimated from bulk $\chi(T)$ measurements.\cite{Honda98} ESR and
specific heat studies revealed a transition to the high-field
phase at $H_c^{\|}=3.4$~T, and $H_c^{\bot}=5.8$~T, extrapolated to
$T\rightarrow 0$, for a magnetic field applied parallel and
perpendicular to the chain-axis,
respectively.\cite{Honda98,Honda99} This anisotropy of critical
field is attributed to single-ion easy-plane magnetic anisotropy
of type $DS_{z}^{2}$. The anisotropy constant was obtained from
bulk susceptibility data: $D/J\approx0.3$. Using the well-known
numerical result $\Delta_z\approx 0.41J+2pD$, $\Delta_{x}\approx
0.41J-pD$, $p\approx 2/3$,\cite{Golinelli93,Meshkov93} one can
thus estimate the Haldane gap energies: $\Delta_{\|}\approx
2.1$~meV and $\Delta_{\bot}\approx 0.54$~meV.

While NDMAP appears to be an ideal model system for neutron
scattering experiments in the high-field phase, additional
characterization, particularly a measurement of inter-chain
interactions and in-plane anisotropy, is required before such a
study can be carried out. Finding the 3D AF zone-center is
especially important, since it is at this wave vector that static
long-range correlations are expected to appear in the high-field
phase. Obviously, inelastic neutron scattering is the most direct
method to obtain this information. In the present paper we report
the results of a zero-field inelastic cold-neutron scattering
study of deuterated NDMAP single-crystal samples, aimed at
extracting this information.

\section{Experimental procedures and results}
Fully deuterated NDMAP single crystal samples were grown from
solution as described in Ref.~\onlinecite{Monfort96}. It was
observed that the crystals tend to shatter when cooled to low
temperature, and even more so when warmed back up.   When wrapped
in aluminum foil the crystals do not fall apart, but the width of
the mosaic distribution increases dramatically with thermal
cycling. Most inelastic neutron scattering data were collected on
a 140~mg single crystal sample that was taken trough the cooling
cycle only twice. The mosaic of the as-grown sample was roughly
$0.3^{\circ}$ and increased to $1.5^{\circ}$ and $3^{\circ}$ after
the first and second cooling, respectively. Inelastic neutron
scattering measurements were performed on the SPINS 3-axis
spectrometer installed on the cold neutron source at the National
Institute of Standards and Technology Center for Neutron Research.
Pyrolitic graphite crystals set for their (002) reflection were
used as monochromator and analyzer. Beam divergences were
approximately $40'-80'-80'-240'$ through the instrument, with a
cooled Be filter between the monochromator and sample. The
measurements were done with a fixed final neutron energy
$E_f=2.8$~meV. The crystal was mounted with either the $a$ or $b$
crystallographic axis vertical, making $(0,k,l)$ and $(h,0,l)$
reciprocal-space planes accessible to measurements. The sample was
cooled to $T=1.4$~K in an "ILL-Orange" cryostat with a 70 mm
diameter sample well.

\begin{figure}
\parbox[b]{3.4in}{
\psfig{file=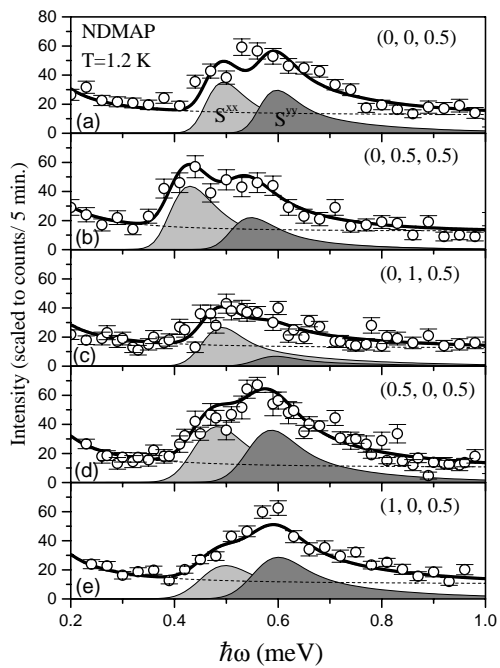,height=4.2in}\caption{Typical constant-$Q$
scans measured at $T=1.2$~K in an NDMAP deuterated single-crystal.
Solid lines are based on global fits to the data, as described in
the text. The shaded areas represent partial contributions of the
$a$-axis and $b$-axis polarized Haldane gap excitations to the
intensity. The dashed line indicates the background level. The
magnetic excitations are resolution-limited and peak shapes are
entirely defined by resolution effects.} \label{exdata}}
\end{figure}

The primary goal of the experiment was to determine the $a$- and
$b$-axis dispersion relations and the polarizations of the two
lower-energy Haldane-gap excitations, relevant for the soft-mode
transition at $H_c$. Most of the data were collected in
constant-$Q$ scans at the 1D AF zone-center $l=0.5$ in the energy
transfer range 0--1~meV. In these dispersion  measurements
$Q$-resolution is particularly important, so a flat analyzer was
used. Typical scans for different momentum transfers perpendicular
to the $c$-axis are shown in Fig.~\ref{exdata}. At all wave
vectors a well-defined peak corresponding to the lower-energy
Haldane excitation doublet is clearly seen around 0.5~meV energy
transfer.

\begin{figure}
\parbox[b]{3.0in}{
\psfig{file=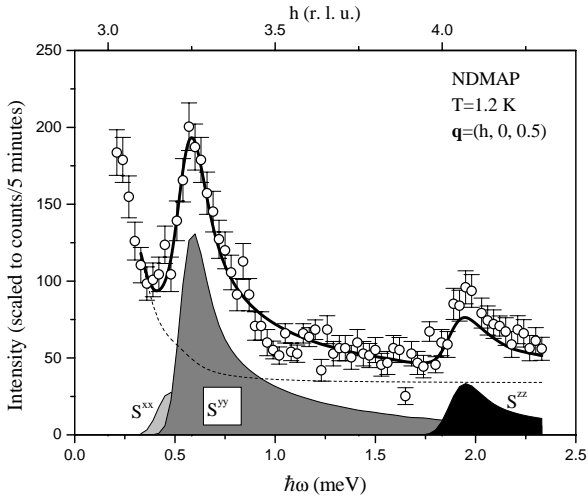,height=3.0in,angle=-90}\caption{ A
constant-$Q_{\|}$ scan measured in NDMAP using a
horizontally-focusing analyzer configuration. Top axis shows the
momentum transfer perpendicular to the chain-axis. Lines are as in
Fig.~\protect\ref{exdata}.} \label{exdata2}}
\end{figure}

To observe the higher-energy Haldane-gap excitation and obtain an
accurate measurement of the anisotropy constant we performed a
constant-$Q$ scan in the range 0--2.4~meV
(Fig.~\ref{exdata2}). To maximize intensity we
used a horizontally-focusing analyzer pointing ${\bf c}^*$ towards the
analyzer to maintain wave vector resolution along the chain. The scattering
angle was varied so the projection of wave vector transfer along the chain was
$0.5 {\bf c}^*$ throughout the scan. In addition to the peak seen in the low-energy scans,
a weaker feature is observed at $\hbar \omega \approx 2$~meV that can be
attributed to the $c$-axis-polarized Haldane gap mode.

\section{Data analysis and discussion}
The measured constant-$Q$ scans were analyzed using a
parameterized model cross section, numerically convoluted with the
Cooper-Nathans 3-axis spectrometer resolution function. Near the
1D AF zone-center the single mode approximation (SMA) for the
dynamic structure factor of isolated Haldane spin chain is known
to work extremely well.\cite{Ma92,Xu96} For each channel of spin
polarization the dynamic structure factor  $S^{(\alpha
\alpha)}(\bbox{q},\omega)$ can be written as:
\begin{eqnarray}
S^{(\alpha
\alpha)}(\bbox{q},\omega)\propto|f(q)|^2\left(1-\frac{\bbox{qe}_{\alpha}}{q^2}\right)\times
\nonumber \\ \times \frac{1-\cos(\bbox{qc})}{2}
\frac{Zv}{\omega_{\alpha,\bbox{q}}}\delta(\hbar \omega- \hbar
\omega_{\alpha,\bbox{q}}),\label{e1}\\
(\hbar\omega_{\alpha,\bbox{q}})^2=\Delta_{\alpha}^2+v^2\sin^2(\bbox{q}\bbox{c}).
\end{eqnarray}
Here $v$ is the spin wave velocity, given by $v\approx
2.49J$.\cite{Sorensen94} The dimension-less constant $Z$ defines
the static staggered susceptibility of a Haldane spin chain:
$\chi_{\pi}=Zv/\Delta$.\footnote{Here we have adopted the notation
used in Ref.~\protect\onlinecite{MZ98-L}} In the above expression
we have included the magnetic form factor $f(q)$ for Ni$^{2+}$ and
the polarization factor
$\left(1-\frac{\bbox{qe}_{\alpha}}{q^2}\right)$. The structure
factor for weakly-coupled chains can be calculated in the Random
Phase Approximation (RPA).\cite{Morra88} The expression for
$S^{(\alpha \alpha)}(\bbox{q},\omega)$ does not change explicitly,
but the excitations acquire dispersion perpendicular to the chain
axis:
\begin{equation}
(\hbar\omega_{\alpha,\bbox{q}})^2=\Delta_{\alpha}^2+v^2\sin^2(\bbox{q}\bbox{c})+ZvJ'(\bbox{q}).\label{e2}
\end{equation}
In this formula $J'(\bbox{q})$ is the Fourier transform of
inter-chain exchange interactions, that we assume to be isotropic
(of Heisenberg type).  According to numerical calculations
$Z\approx 1.26$.\cite{Sorensen94} The form of $J'$ can be guessed
by looking at the crystal structure. The smallest inter-chain
Ni-Ni distance (8.705\AA) is along the $b$ crystallographic axis.
We shall denote the corresponding exchange constant as $J_y$. The
next-smallest inter-chain Ni-Ni distance (10.478\AA) is along the
(0.5,0.5,0.5) direction. This interaction, however, is frustrated
by in-chain AF interactions (any site in one chain couples to two
consecutive sites in another chain), and is thus irrelevant,
within the RPA, at momentum transfers $\bbox{qc}\approx \pi$.
Finally, the third-smallest inter-chain distance (18.046\AA) is
along $a$. The corresponding coupling constant $J_x$ is expected
to be very small, due to the large ``bond'' length. The Fourier
transform of inter-chain interactions can thus be written as:
\begin{equation}
J'(\bbox{q})=2J_x\cos(\bbox{qa})+2J_y\cos(\bbox{qb}).\label{e3}
\end{equation}

The chain-axis exchange constant $J$ is not very important for our
measurements: for scans collected at the 1D AF zone-center it only
slightly influences the peak shapes, due to a non-zero
$Q$-resolution along the chain-axis. We can therefore safely use
the value previously obtained from susceptibility measurements.
The relevant parameters of the model are thus the three gap
energies $\Delta_x$, $\Delta_y$ and $\Delta_z$, two inter-chain
exchange constants $J_x$ and $J_y$ and an overall intensity
prefactor. Three additional parameters were used to describe the
background: an energy-independent component, and the intensity and
width for a Lorenzian profile centered at $\hbar \omega=0$ to
account for incompletely resolved elastic scattering.

\begin{figure}
\parbox[b]{3.4in}{
\psfig{file=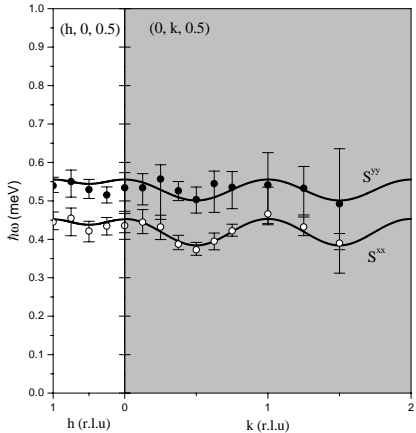,height=3.0in,angle=-90}\caption{ Transverse
dispersion of lower-energy Haldane-gap excitations in NDMAP, as
determined in global fits to constant-$Q$ scans (solid lines), as
described in the text. The data points represent excitation
energies obtained in fits to individual scans.} \label{disp}}
\end{figure}

The difference between $\Delta_x$ and $\Delta_y$, usually a result
of  in-plane magnetic anisotropy of type $E(S_x^2-S_y^2)$, is too
small to observe two and separate corresponding peaks in any
single constant-$Q$ scan. Distinct polarization factors for the
two branches however allow us to extract both parameters in a
global fit to the data measured at different momentum transfers
perpendicular to the chain-axis. As a first step, we
simultaneously analyzed the data collected in $(0,k,l)$
reciprocal-space plane (11 scans with a total of 382 data points).
A very good global fit using Eqs.~\ref{e1} and \ref{e2} is
obtained with a residual $\chi^2=1.6$. The solid lines in
Fig.~\ref{exdata}a--c were calculated from the globally optimized
parameter set. The shaded areas represent the contribution of each
mode, and the dashed lines show the background level. Figure
\ref{disp} (right) shows the obtained $b$-axis dispersion relation
(solid line), that has a minimum at $k=0.5$. Symbols in this
figure represent the excitation energies obtained in fits to
individual scans. A similar global fit (4 scans, 149 data points,
$\chi^2=1.7$) was applied to all constant-$Q$ scans measured in
the $(h,0,l)$ scattering plane (solid lines in Fig.~\ref{exdata}d
and e). In this case the excitation energies at the zone-boundary
$(0,0,0.5)$ were fixed to the values obtained from the
$(0,k,l)$-plane global fit. Only the $a$-axis exchange constant
$J_x$ was refined. Dispersion of magnetic excitations along this
direction is barely detectable. The dispersion relation obtained
from our fits is nontheless plotted in a solid line in
Fig.~\ref{disp} (left). The fitting analysis suggests a shallow
minimum at $h=0.5$. From Eq.~\ref{e3} we can thus guess that the
global minimum of the 3D dispersion (3D AF zone-center) is located
at $(0.5,0.5,0.5)$. The energy gap for $c$-axis polarized
excitations was determined by fitting the model cross section to
the wide-range constant-$Q$ scan measured with the
horizontally-focusing configuration (Fig.~\ref{exdata2}, solid
line). In this procedure the values of $\Delta_x$ and $\Delta_y$
were fixed. The parameters determined by the analysis described
above are as follows: $\Delta_x=0.42(0.03)$~meV,
$\Delta_y=0.52(0.06)$~meV, $\Delta_z=1.86(0.1)$~meV, $J_y=1.8
(0.4) \cdot10^{-3}$~meV, and $J_x=3.5(3.0) \cdot10^{-4}$~meV.

The in-chain exchange constant $J$ and anisotropy parameter $D$
are readily obtained from the measured gap
energies:\cite{Golinelli93,Meshkov93} $J\approx
0.81(\Delta_x+\Delta_y+\Delta_z)=2.28$~meV, $D\approx
\frac{1}{4}(2\Delta_z-\Delta_x-\Delta_y)=0.70$~meV, and
$D/J=0.30$, which is consistent with the bulk susceptibility
result of Ref.~\onlinecite{Honda98} The relative strength of
inter-chain interactions are $J_y/J\approx 7 \cdot 10^{-4}$ and
$J_x/J\approx 1.3 \cdot 10^{-4}$. These ratios are very similar to
those found in NENP.\cite{NENP}

\section{Summary}
The small value of critical fields in NDMAP, compared to those in
NENP, are a result of smaller in-chain exchange interactions and
somewhat larger easy-plane magnetic anisotropy. The lowest-energy
excitation in NDMAP is polarized along the crystallograhic
$a$--axis. The 3D magnetic zone-center appears to be
$(0.5,0.5,0.5)$. Future experimental studies of the high-field
phase should thus concentrate on this region of reciprocal space.

\acknowledgements We would like to thank S. M. Shapiro for
illuminating discussions, S.-H. Lee for his assistance with the
measurements at NIST, and R. Rothe for technical support at BNL.
Work at Brookhaven National Laboratory was carried out under
Contract No. DE-AC02-98CH10886, Division of Material Science,
U.S.\ Department of Energy. Work at JHU was supported by the NSF
through DMR-9801742. This work used instrumentation supported by
NIST and the NSF through DMR-9423101. Work at RIKEN was supported
in part by a Grant-in-Aid for Scientific Research from the
Japanese Ministry of Education, Science, Sports and Culture.

\end{document}